\documentclass[11pt,a4paper,english]{article}
\usepackage{jheppub}
\usepackage{amsfonts,amsthm,babel,multicol,fancyhdr, slashed,eucal}
\usepackage{mathrsfs}
\usepackage{simplewick} 
\usepackage{xcolor}
\usepackage{appendix}

\usepackage[utf8]{inputenc}

\graphicspath{{home/}}

\numberwithin{equation}{section}

\newcommand{\bo}{\boldsymbol}
\newcommand{\smallfrac}[2]{{\textstyle\frac{#1}{#2}}}
\newcommand{\BE}{\begin{equation}}
\newcommand{\EE}{\end{equation}}
\newcommand{\mrm}{\mathrm}

\newcommand{\dd}{\mathrm{d}}

\newcommand{\me}{\mathrm{e}}
\newcommand{\mcal}{\mathcal}
\newcommand{\mn}{\mathnormal}
\newcommand{\del}{\partial}

\newcommand{\con}{\mrm{c}}
\newcommand{\mLam}{\mn\Lambda}

\newcommand{\mSig}{\mn\Sigma}

\newcommand{\mPhi}{\mn\Phi}
\newcommand{\mDelt}{\mn\Delta}
\newcommand{\fdelphi}{ \frac{\delta}{\delta \phi} }
\newcommand{\fdelphiO}[1]{ \frac{\delta #1}{\delta \phi} }

\title{Stochastic Renormalization Group and Gradient Flow}

\author{Andrea Carosso}

\affiliation{Department of Physics, University of Colorado, \\
Boulder, Colorado 80309, USA}

\emailAdd{andrea.carosso@colorado.edu}

\abstract{A non-perturbative and continuous definition of RG transformations as stochastic processes is proposed, inspired by the observation that the functional RG equations for effective Boltzmann factors may be interpreted as Fokker-Planck equations. The result implies a new approach to Monte Carlo RG that is amenable to lattice simulation. Long-distance correlations of the effective theory are shown to approach gradient-flowed correlations, which are simpler to measure. The Markov property of the stochastic RG transformation implies an RG scaling formula which allows for the measurement of anomalous dimensions when transcribed into gradient flow expectation values.}

\arxivnumber{1904.13057}

\begin{document}

\maketitle

\flushbottom

\section{Introduction}

There has been much recent work on the subject of how gradient flow (GF) \cite{Narayanan:2006rf, Luscher:2009eq} is related to the renormalization group (RG), due to the characteristic smoothing properties of GF resembling the high-mode elimination aspect of RG. Inspired by this resemblance, work has been done on the lattice to use gradient flow to define a continuous blocking transformation with which anomalous dimensions have been computed \cite{Carosso:2018bmz,Carosso:2018rep}, with encouraging results. On the analytic side, work has been done \cite{Makino:2018rys, Abe:2018zdc, Sonoda:2019ibh} connecting GF to the framework of functional (or exact) RG (FRG), which is a method by which one defines RG transformations nonperturbatively and continuously in the context of continuum field theory \cite{Wegner:1972ih, Wilson:1973jj, Polchinski:1983gv, Wetterich:1989xg}.\footnote{See \cite{Kopietz:2010zz, Rosten:2010vm} for exemplary reviews of the subject.} In particular, it has been noted that certain definitions of a GF effective action lead to a kind of Langevin equation \cite{Abe:2018zdc}, and most recently, that the connected $n-$point functions of a particular FRG effective theory are equal to the GF observables up to proportionality \cite{Sonoda:2019ibh}.

In this work, we will contribute to these discussions by demonstrating the equivalence of certain kinds of FRG transformations with a class of stochastic (Markov) processes on field space.\footnote{See \cite{Pavliotis:2014} for a mathematician's introduction to stochastic processes, \cite{ZinnJustin:2002ru} for a physicist's introduction, and \cite{Damgaard:1987rr, ZinnJustin:2002ru} for an introduction to their field-theoretical generalization in the context of stochastic quantization.} It has been noted before \cite{Gaite:2000jv, Pawlowski:2017rhn, ZinnJustin:2007zz} that the functional RG equations for effective actions, when written in terms of effective Boltzmann weights, are of the form of a Fokker-Planck (FP) equation, whose solution is therefore a probability distribution over effective fields. Taking this observation seriously, and recalling that Fokker-Planck distributions can be thought of as being generated by a Langevin equation on the degrees of freedom appearing in the FP distribution, one may ask what kinds of Langevin equation generate the FRG effective actions. In what follows, we will define an RG transformation by a particularly simple (linear) choice of Langevin equation, and show by direct calculation that the transition functions resemble the constraint functionals found in the literature of FRG. The effective action for the specific case of $\phi^4$ theory in three dimensions will then be discussed, and the existence of a nontrivial IR fixed point will be checked to 1-loop order in perturbation theory. It will therefore become apparent that although the stationary distribution of the FP equation would be expected to be gaussian, a simple rescaling of variables allows for an interacting fixed point solution.\footnote{This may not be surprising from the RG perspective, of course, but it may be unexpected from the standpoint of stochastic processes, where the stationary distributions of the Smoluchowski equation are expected to involve the potential whose gradient appears in the Langevin equation\cite{Pavliotis:2014}.}

The equivalence we discuss here is a formulation of the Monte Carlo Renormalization Group (MCRG) principle for FRG. Recall that the kind of MCRG discussed by Swendsen \cite{Swendsen:1979gn} in the 1980's provided a prescription for computing observables in an effective theory by computing \emph{blocked} observables in a bare theory, that is, without having to know the effective action. A similar property will be found for the stochastic RG transformation, namely, that effective observables may be computed from the stochastic observables generated by the Langevin equation, whose initial condition is the bare field. The MCRG property will be valid for both lattice and continuum theories alike, thereby suggesting the possibility of computing general observables in an effective theory on the lattice by integrating a Langevin equation on top of the ensemble generated in the MCMC simulation of the corresponding bare theory. 

The relationship to gradient flow will then follow from an observation made by Wilson and Kogut \cite{Wilson:1973jj}, and recently connected to gradient flow by Sonoda and Suzuki \cite{Sonoda:2019ibh}. In the context of the stochastic RG transformation, it follows from the MCRG equivalence that the connected expectation values of an FRG effective theory are equal to gradient-flowed expectations up to additive corrections that depend on the choice of Langevin equation, and which decay exponentially at large distances. This relationship implies that the measurement of gradient-flowed quantities is sufficient for the determination of long-distance critical properties of the theory, in much the same way as spin-blocked observables at large distances, avoiding the necessity of performing a full Langevin equation simulation.

A virtue of the characterization of FRG in terms of stochastic processes is that the properties of the Langevin equation guarantee certain traditionally desired properties of an FRG transformation: a composition law for repeated applications of the transformation, a correct initial condition for the effective theory, and a natural characterization of the fixed points of the transformation. Moreover, that the effective Boltzmann factor satisfies a FP equation implies that observables of the effective theory satisfy a similar equation involving the generator of the Markov process. An analysis of this equation for discrete, small time steps leads to the stochastic RG instantiation of usual RG scaling laws of correlations of the fundamental field, as well as of composite operators built from it. In particular, by virtue of the stochastic MCRG equivalence, one is led to correlator ratio formulae of the sort described in \cite{Carosso:2018bmz,Carosso:2018rep}, implying a method for measuring anomalous dimensions of scaling operators close to a critical fixed point.

\section{Stochastic processes and FRG}

Here we discuss the general framework for stochastic RG. The RG transformation will be defined by a Langevin equation on the degrees of freedom of a field theory. The simplicity of the equation will allow for an explicit calculation of the probability distribution which it generates, and the functional form of the distribution will entail an equivalence to conventional FRG transformations. A brief consideration of the observables generated by the stochastic process will lead to the MCRG equivalence between the effective theory and the stochastic observables. Lastly, we will comment on the pitfalls of a seemingly simpler definition of the effective theory.

\paragraph{The Langevin equation.}

We will define an RG transformation by a stochastic process $\phi_t$ on field space over $\mathbb{R}^d$ determined by a Langevin equation (LE) of the form
\BE \label{LE_momspace}
\del_t \phi_t(p) = -\omega(p) \phi_t(p) + \eta_t(p), \quad \phi_0(p) = \varphi(p),
\EE
where $\omega(p)$ is positive for $\|p\|>0$ and $\omega(0) \geq 0$, e.g. $\omega(p) = p^2$, where $p^2 := \| p \|^2$.\footnote{Of course, the realm of stochastic quantization \cite{Damgaard:1987rr} deals with writing field theory expectation values as equilibrium limits of a stochastic process on field space. Here, however, the bare theory is kept as a traditional field theory, and the stochasticity applies to the RG transformation only.} The ``time'' $t$ in this equation does not denote physical time, but rather an ``RG time'' which we will call \emph{flow time}, or simply \emph{time}. The noise $\eta_t(p)$ is chosen to be gaussian-distributed according to the measure
\BE \label{noise_distribution}
\dd \mu_{0} (\eta) := \mrm{c}(\mLam_0, \Omega) \exp\Big[ - \frac{1}{2\Omega} \int_I \dd t \; (\eta_t, K_{0}^{-1}\eta_t)\Big] \mathscr{D} \eta,
\EE
where the notation $(\phi, M \chi)$ denotes a quadratic form, written variously as\footnote{We abbreviate $\int_x = \int_{\mathbb{R}^d} \dd^d x$ and $\int_p = \int_{\mathbb{R}^d} \dd^d p / (2\pi)^d$ when no confusion arises.}
\BE
(\phi, M \chi) =  \int_{xy} \phi (x) M(x,y) \chi(y) = \int_{pk} \phi(p) M(p,k) \chi_t(k).
\EE
The cutoff function $K_{0}(p)$ suppresses noise momentum modes greater than $\mLam_0$, e.g. $K_{0}(p) = \me^{-p^2/\mLam_0^2}$ under Schwinger regularization.\footnote{The LE and measure $\dd \mu_0$ can easily be written for a lattice theory, in which case the cutoff function $K_0$ is not necessary, as the lattice naturally regulates the noise at the bare scale.} Expectation values with respect to the noise distribution of functions $O(\eta)$ are defined by
\BE
\mathbb{E}_{\mu_0}[O(\eta)] := \int \! O(\eta) \dd \mu_0 (\eta).
\EE
The first two moments of $\mu_{0}$ are then
\BE
\mathbb{E}_{\mu_0}[ \eta_t(p) ] = 0, \quad  \mathbb{E}_{\mu_0}[ \eta_t(p) \eta_s(k) ]  = 2\pi \Omega \; \delta(t-s) \; (2\pi)^d \delta(p+k) K_{0}(k).
\EE
Later we will take the initial condition $\phi_0 = \varphi$ to be distributed according to a measure $\dd \rho_0 (\varphi)$ corresponding to the bare theory of interest, the cutoff of which is chosen to be $\mLam_0$. Hence, the cutoff for the noise is chosen to match the cutoff of the bare theory.

Turning back to eq. (\ref{noise_distribution}), the constant $\mrm{c}(\mLam_0, \Omega)$ is chosen to normalize $\dd \mu_{0}$ to unity, $\Omega$ is the (dimensionless) variance of the noise, and $I \subset \mathbb{R}$ is an arbitrary time interval large enough to include all desired times. In position space, the Langevin equation takes the form of a stochastic heat equation
\BE
\del_t \phi_t(x) =  - (\omega \phi_t)(x) + \eta_t(x), \quad \phi_0(x) = \varphi(x).
\EE
For the case $\omega(p) = p^2$, one has $(\omega \phi)(x) = -\Delta \phi (x) = -\del_\mu \del_\mu \phi(x)$. In position space, therefore, we see that the equation becomes a stochastic partial differential equation.

The form of the momentum space equation above is a simple field-theoretic generalization of the well-known Ornstein-Uhlenbeck process (i.e. damped Brownian motion) $q_t$ with Langevin equation and solution \cite{ZinnJustin:2002ru}, respectively,
\BE
\dot q_t = - \omega q_t + \eta_t, \quad q_t = \me^{-\omega t} q_0 + \int_0^t \dd s \; \me^{-\omega(t-s)} \eta_s,
\EE
where $\eta_t$ is gaussian white noise, so it is quite simple to find the solution. One treats the noise term like a non-homogeneous part of the equation, finding
\BE
\phi_t(p) = f_t(p) \varphi(p) + \int_0^t \dd s \; f_{t-s}(p) \eta_s(p),
\EE
where $f_t(p)$ is a generalized momentum space heat kernel of the form
\BE
f_t(p) = \me^{-\omega(p)t}, \quad f_t(z) = \int_p \me^{i p \cdot z} f_t(p).
\EE
In position space, one finds
\BE
\phi_t(x) = (f_t \varphi)(x) + \int_0^t \dd s \; (f_{t-s}\eta_s)(x).
\EE
We will sometimes denote the solution's dependence on initial condition and noise by $\phi_t[\varphi; \eta]$. The first term on the right-hand side implies that the mean of $\phi_t(x)$ satisfies the free gradient flow equation, i.e. ``heat'' equation, corresponding to the differential operator $\omega$.

\paragraph{The Fokker-Planck distribution.}

With the explicit solution in-hand, one can compute the probability distribution of fields $\phi$ at time $t$ given $\varphi$ at $t=0$. We say that the Langevin equation \emph{generates} a Fokker-Planck (FP) distribution $P(\phi,t;\varphi,0)$ defined by
\BE
P(\phi,t;\varphi,0) := \mathbb{E}_{\mu_0} \big[ \delta(\phi - \phi_t[\varphi;\eta])\big] = \int \mathscr{D} \lambda \; \mathbb{E}_{\mu_0} \big[ \me^{i(\lambda,\phi-\phi_t [\varphi; \eta])}\big],
\EE
From the definition of noise expectations, we then find
\BE
P(\phi, t; \varphi, 0) =  \mrm{c}(\mLam_0, \Omega) \int \! \mathscr{D} \lambda \int \! \mathscr{D} \eta \; \exp\Big[i(\lambda, \phi - \phi_t[\varphi; \eta]) -\frac{1}{2\Omega} \int_I \dd s (\eta_s, K^{-1}_{0} \eta_s)\Big].
\EE
Substituting in the explicit solution for $\phi_t$, the integrand becomes
\begin{align}
\exp&\Big[i(\lambda, \phi - f_t \varphi)  - i \int_0^t \dd s\; (\lambda, f_{t-s} \eta_s) - \frac{1}{2\Omega} \int_I \dd s \; (\eta_s, K_{0}^{-1} \eta_s)\Big] \nonumber \\
& = C \exp\Big[i(\lambda, \phi - f_t \varphi)  - \int_0^t \dd s\; \Big( i (\lambda, f_{t-s} \eta_s) + \frac{1}{2\Omega} (\eta_s, K_{0}^{-1} \eta_s) \Big) \Big],
\end{align}
the constant $C$ involving times $s > t$, which divide out of any noise averages and will now be dropped. The noise integral over relevant $\eta_t$'s is a standard gaussian integral, which yields
\BE
P(\phi, t; \varphi, 0) = \int \! \mathscr{D}\lambda \; \exp\Big[i(\lambda, \phi - f_t \varphi) - \frac{\Omega}{2} \int_0^t \dd s \; (f_{t-s}^\top \lambda, K_{0} f_{t-s}^\top
\lambda)\Big].
\EE
Next, note that the $s-$integral (which does not care about $\lambda$ or $K_{0}$) produces a kernel
\BE \label{Adef}
A_t := \Omega \int_0^t \dd s \; f_{t-s} K_{0} f_{t-s}^\top,
\EE
which in momentum space is given by a diagonal matrix, 
\BE \label{A_momspace}
A_t(p,k) = \Omega (2\pi)^d \delta(p+k)  K_{0}(p) \; \frac{1-\me^{-2 \omega(p) t}}{2\omega(p)}.
\EE
We will sometimes denote $A^{-1}_t$ by $B_t$; the inverse exists by virtue of the restrictions on $\omega(p)$. The remaining $\lambda-$integral is also gaussian, and evaluates to
\BE \label{FP_dist}
P(\phi, t; \varphi, 0) = \big[\det 2 \pi B_t \big]^{\frac{1}{2}} \exp\Big[- \frac{1}{2} \big(\phi - f_t\varphi, B_t (\phi - f_t \varphi)\big)\Big].
\EE
Here we note that similar functionals to this one are common to several formulations of FRG \cite{Wilson:1973jj, Wetterich:1989xg, Igarashi:2009tj}. We will call such a functional a \emph{gaussian constraint functional} or a \emph{transition function} (when emphasizing it's probabilistic interpretation). In momentum space, the exponent is explicitly
\BE \label{constraint_functional}
- \frac{1}{2} \int_p \; \frac{2\omega(p) K_{0}^{-1}(p)}{1-\me^{-2\omega(p)t}} \big(\phi(p) - \me^{-\omega(p)t} \varphi(p)\big)\big(\phi(-p) - \me^{-\omega(p)t} \varphi(-p)\big).
\EE
One observes that the mean of the field $\phi$ is set to the flowed field $f_t(p) \varphi(p)$ within a functional variance determined by $A_t(p)$. Thus the effect of the stochastic RG transformation is to produce a low-mode fluctuating field, in the sense that the mean value of modes of $\phi$ with $p^2 \gg 1/t$ are exponentially suppressed. This suggests that the effective cutoff of the resulting theory is roughly $\mLam_t \sim 1/\sqrt{t}$ ; a more precise identification will be made later. 

For reasons explained in the next subsection, we may write the transition function as $P_t(\phi,\varphi)$ rather than $P(\phi,t;\varphi,0)$, and we will sometimes suppress the initial condition by writing $P_t(\phi)$. The transition function is a Green function for the Fokker-Planck equation\footnote{The tensor products are in the space of position or momentum indices.}
\begin{align} \label{FP_transfn}
& \frac{\del P_t(\phi)}{\del t} = \mrm{tr} \; \fdelphi \otimes \Big( \frac{1}{2} \mSig(\phi,t) \fdelphiO{P_t(\phi)} + \mathscr{B}(\phi,t) P_t(\phi) \Big), \nonumber \\ 
& \lim_{t \to 0} P(\phi,t; \varphi,0) = \delta(\phi - \varphi),
\end{align}
where the drift $\mathscr{B}$ and diffusion matrix $\mSig$ are defined by \cite{Pavliotis:2014}
\begin{align}
\mathscr{B}(\phi,t) &= \lim_{t' \to t} \frac{1}{t'-t} \int_{\phi'} (\phi' - \phi) P(\phi',t';\phi,t), \\
\mSig(\phi,t) &= \lim_{t' \to t} \frac{1}{t'-t} \int_{\phi'} (\phi' - \phi) \otimes (\phi' - \phi) P(\phi',t';\phi,t).
\end{align}
With the explicit solution eq. (\ref{FP_dist}), we compute
\begin{align}
\mathscr{B}(\phi,t) &= \omega \phi, \\
\mSig(\phi,t) &= \Omega K_{0}.
\end{align}
If the initial condition $\varphi$ is distributed according to a measure $\dd \rho_0(\varphi) = \me^{-S_0(\varphi)} \mathscr{D} \varphi$ corresponding to a bare theory, then the effective distribution
\BE
\rho_t(\phi) = \int \! P(\phi,t;\varphi,0) \; \dd \rho_0(\varphi)
\EE
also satisfies the FP equation, with initial condition $\rho_0(\varphi)$. For the specific choice eq. (\ref{LE_momspace}) of Langevin equation above, we find
\BE \label{FP_EFT}
\frac{\del \rho_t(\phi)}{\del t} = \frac{1}{2} \mrm{tr} \; K_0 \; \fdelphi \otimes \Big( \fdelphiO{\rho_t(\phi)} + 2 \mDelt_{\omega}^{-1} \phi \; \rho_t(\phi) \Big),
\EE
where we have set $\Omega = 1$, and $\mDelt_{\omega} = K_{0} \omega^{-1}$ is a \emph{regularized} propagator.

The drift term $(\omega \phi_t)(x)$ may be regarded as the functional derivative of what we might call a ``flow action''
\BE
\hat S(\phi) = \frac{1}{2} (\phi, \omega \phi), \quad \del_t \phi_t = - \fdelphiO{\hat S}(\phi_t) + \eta_t,
\EE
in which case one would have $\mathscr{B} = \fdelphiO{\hat S}$.
For arbitrary choices of $\hat S$, the Langevin equation may become nonlinear and the (still linear) FP equation generalizes to 
\BE \label{FP_seed}
\frac{\del \rho_t(\phi)}{\del t} = \frac{1}{2} \mrm{tr} \; K_{0} \fdelphi \otimes \Big( \fdelphiO{\rho_t(\phi)} + 2 K^{-1}_{0} \fdelphiO{\hat S(\phi)} \rho_t(\phi) \Big).
\EE
Similar equations are common in the FRG literature, where they are sometimes referred to as linear FRG equations or generalized heat equations \cite{Rosten:2010vm}. Of course, by writing $\rho_t = \me^{-S_t}$ and letting $\dd \rho_0(\varphi) = \me^{-S_0(\varphi)} \mathscr{D}\varphi$, one recovers functional PDEs for the effective action $S_t(\phi)$ given some bare action $S_0(\varphi)$, similar to the Polchinski equation.

There are many possibilities for how to generalize the scheme presented above. First, one could choose a different distribution for the noise, perhaps even a non-gaussian one. Second, one could generalize the flow action to be arbitrarily complicated in $\phi$, thereby making the Langevin equation non-linear, but these will generate FP distributions which are more difficult to calculate, and it is unclear what is gained by doing so, especially because we will find that the linear equation is sufficient to define a proper RG transformation. For theories whose field variables are in compact spaces, or theories with local symmetries, however, one will need non-linear LEs to ensure that the flow preserves the symmetry; such equations will likely resemble those found in the context of stochastic quantization \cite{Damgaard:1987rr, Batrouni:1985jn}.

\paragraph{Further properties and MCRG.} Although the distribution above has the same form as the constraint functionals found in the FRG literature, a notable difference here is that the kernel $B_t$ is \emph{determined} by the associated Langevin equation having a fixed relation to $\omega$, the choice of drift. The initial condition for the transition function, $P_0(\phi,\varphi) = \delta(\phi - \varphi)$, is guaranteed by the fact that it is generated by a LE with initial condition $\varphi$. As a distribution, it is furthermore normalized such that for all $t \geq 0$,
\BE
\int \! \mathscr{D} \phi \; P_t(\phi,\varphi) = 1,
\EE
and in particular, the integral is independent of the field $\varphi$. These conditions allow one to define the effective theory in a more conventional way by inserting unity into the partition function $Z$ of the bare theory as
\BE
Z = \int \! \dd \rho_0(\varphi)  = \int \! \mathscr{D} \varphi \! \int \! \mathscr{D} \phi \; P_t(\phi,\varphi) \; \me^{-S_0(\varphi)},
\EE
thereby defining a Boltzmann weight of effective (low-mode) fields
\BE \label{eff_action}
\dd \rho_t(\phi) = \frac{1}{Z} \; \me^{-S_t(\phi)} \mathscr{D} \phi , \qquad \me^{-S_t(\phi)} := \int \! \mathscr{D} \varphi \; P_t(\phi,\varphi) \; \me^{-S_0(\varphi)},
\EE
and the partition function remains invariant.

The stochastic process generated by a Langevin equation is a Markov process, so that future states depend only on the present state, so long as the noise at different times are uncorrelated. This kind of feature was desirable at least in Wilson's philosophy of RG, where any particular blocking step could be carried out by knowing only the previous step. In terms of the abstract distribution $P$ this implies, $\forall \; t > s \geq 0$,
\BE
P(t,0) = P(t, s)P(s, 0), \quad \mrm{or} \quad P(\phi, t; \varphi, 0) = \int \! \mathscr{D} \chi \; P(\phi, t; \chi, s) P(\chi, s; \varphi, 0).
\EE
By considering time-homogeneous Langevin equations (i.e. no explicit $t-$dependence in the LE or the noise variance), the transition function depends only on the difference $t-s$, and we can write $P(\phi, t; \chi, s) = P_{t-s}(\phi, \chi)$.\footnote{The noise variance can be chosen to depend on time, but this spoils the convenience of time-homogeneity.} This entails that the set $\{P_t : t\geq 0\}$ form a semigroup of operators and may be written in terms of a generator $\mcal{L}$ as $P_t = \me^{t\mcal{L}}$ \cite{Pavliotis:2014}. We will discuss $\mcal{L}$ in the last section. For now we simply note that $\mcal{L}$ is the adjoint of the functional differential operator appearing in the FP equation.

Next, consider the usual definition of the expectation value of an operator $\mcal{O}$ in the effective theory,
\BE
\langle \mcal{O}(\phi) \rangle_{S_t} := \frac{1}{Z} \int \! \mathscr{D} \phi \; \mcal{O}(\phi) \; \me^{- S_t(\phi)}.
\EE
By inserting the definition eq. (\ref{eff_action}), and noting that
\BE
\int \! \mathscr{D} \phi \; \mcal{O}(\phi) P(\phi, t; \varphi, 0) = \mathbb{E}_{\mu_0} \big[ \mcal{O}(\phi_t[\varphi;\eta]) \big],
\EE
where $\phi_t[\varphi; \eta]$ denotes the solution of the LE, one readily obtains the equality
\BE \label{Equivalence}
\big\langle \mcal{O}(\phi) \big\rangle_{S_t} = \big\langle \mathbb{E}_{\mu_0} \big[\mcal{O}(\phi_t[\varphi; \eta]) \big] \big\rangle_{S_0}.
\EE
This formula states the equivalence of a low-mode FRG effective theory (in the Polchinski sense, i.e., no rescaling is yet performed) and a double expectation value over the bare fields and the random noise. Since the right-hand side may be calculated without knowledge of the effective action, it further constitutes a generalization of MCRG to FRG for all observables. Notice that there are just as many degrees of freedom $\phi$ as there are $\varphi$ (this is especially clear on the lattice).

In the next section, we will explore various properties of the effective action $S_t(\phi)$ defined above. First, however, one might wonder why the noise average is necessary in eq. (\ref{Equivalence}), when compared with the corresponding statement for a spin-blocked theory \cite{Swendsen:1979gn},
\BE
\big\langle \mcal{O}(\phi) \big\rangle_{S_b} = \langle \mcal{O}(B_b\varphi) \rangle_{S_0},
\EE
where $B_b$ denotes the blocking operator. This is perhaps clarified by the fact that when spin-blocking, there are fewer blocked spins than bare spins, so the blocked expectation values really involve an integration over ``extra'' degrees of freedom, from the perspective of the effective theory; here the role is played by noise. If one were to choose a blocked lattice of the same size as the original, so that the bare Boltzmann factor were integrated against a delta functional over the whole lattice, the resulting blocked action would be trivial, namely, $S_0(B^{-1}_b \phi)$. Likewise in the continuum, it has long been assumed \cite{Wetterich:1989xg} that a pure $\delta-$function constraint functional is not sufficient to define a non-trivial Wilson action for continuum FRG. Let us elaborate on this. One might have wanted to define the Wilson action through
\BE \label{delta_constraint}
\me^{-S_t(\phi)} = \int \! \mathscr{D} \varphi \; \delta(\phi - f_t \varphi) \; \me^{-S_0(\varphi)},
\EE
where $f_t\varphi$ is the solution of a gradient flow equation such as
\BE
\del_t \phi_t(x) = \Delta \phi_t(x),
\EE
or some generalization thereof. The problem with this definition is that it generates a trivial effective action, in the sense to be described. In momentum space, the solution is simply $(f_t \varphi)(p) = \me^{-p^2 t} \varphi(p)$, so one can do a linear change of variables in eq. (\ref{delta_constraint}) and compute
\BE
S_t(\phi) = -\mrm{tr} \ln f_t + S_0(f_t^{-1} \phi).
\EE
Hence, the couplings of the new action are exactly computable, and because their dependence on $t$ is trivially determined by how many powers of $\phi$ and $p^2$ appear in each term, without involving any loop corrections, one verifies that the resulting ``Wilson action'' is not acceptable.

Lastly, we remark that the inadequacy of the definition eq. (\ref{delta_constraint}) \emph{does not} mean that the observables computed from gradient-flowed fields are not useful for studying certain RG properties of the system. At the end of the next section, in particular, we will describe how gradient-flowed observables are sufficient for studying the \emph{long-distance} properties of the theory.

\section{The effective theory} In what follows, the nontriviality of the effective action determined by the stochastic RG transformation will be discussed for the example case of $\phi^4_3$ theory. We will show that by a rescaling of variables, the existence of an interacting IR fixed point of the transformation becomes possible. From the point of view of stochastic processes, the result implies that, by merely rescaling degrees of freedom appropriately, the stationary solutions of the Fokker-Planck equation may be non-gaussian even though the Langevin equation is linear. Lastly, the correlation functions of the effective theory will be related to gradient-flowed correlations.

\paragraph{The effective action.} That the EFT defined by a gaussian constraint functional for $\Omega > 0$ is nontrivial can be understood as follows. One may insert the expression eq. (\ref{FP_dist}) for the transition function into the definition of the effective action eq. (\ref{eff_action}), and then expand out the exponent of $P_t(\phi,\varphi)$; the part proportional to $\varphi^2$ modifies the bare theory propagator, and the part linear in $\varphi$ acts as a source term with $J = f_t B_t \phi$. The remaining $\phi^2$ term contributes to the $\phi$ propagator. The result is a relation between effective and bare actions:\footnote{In a sense, this constitutes an exact solution to the FP equation, giving the finite$-t$ distribution $\rho_t$ in terms of the cumulants of $\rho_0$.}
\BE
S_t(\phi) = F_t + \frac{1}{2} (\phi, B_t \phi) - W_{0}^{(t)}(B_t f_t \phi),
\EE
where $F_t$ is due to the normalization of $P_t(\phi,\varphi)$, and $W^{(t)}_0(J) = \ln \langle \me^{(J,\phi)} \rangle_{S_0^{(t)}}$ is the generator of connected Green functions for the bare theory $S_0$ with a modified $t-$dependent inverse propagator
\BE
[\mDelt^{(t)}_{0}]^{-1} := \mDelt_{0}^{-1} + h_t, \quad h_t := f_t B_t f^\top_t.
\EE
Expanding the generator term in $\phi$ yields a formula which allows for the systematic computation of effective vertices,
\BE
W_0^{(t)}(B_t f_t \phi) = \sum_{n=0}^\infty \frac{1}{n!} \int_{\bo p} [W_0^{(t)}]^{(n)}(\bo p) (B_t f_t \phi)(p_1) \cdots (B_t f_t \phi)(p_n),
\EE
where the set of $n$ momenta $p_i$ is denoted by $\bo p$. It is then apparent that the effective action for any finite $t$ is indeed non-trivial, since the vertices of $S_t$ contain the dynamics of the bare theory via the $[W^{(t)}_0]^{(n)}$. 

The scale $\mLam_t$ of the effective theory may be determined by looking at the effective 2-point function at tree level, after isolating the quadratic part of $S_t(\phi)$:
\BE
\langle \phi(p_1) \phi(p_2) \rangle_{S_t}^\mrm{tree} = (2\pi)^d\delta(p_\mrm{tot}) \Big[ A_t(p) + f_t^2(p) \mDelt_0(p) \Big] = (2\pi)^d \delta(p_\mrm{tot}) \Big[ A_t(p) + \frac{\me^{- p^2(a_0^2 + 2 t)}}{p^2 + m_0^2} \Big],
\EE
where the inverse cutoff $a_0 = \mLam_0^{-1}$ has been used, and we recall that $A_t$ is given by eq. (\ref{A_momspace}). In position space, the first term decays rapidly at large distances with respect to the first. The second term is a Schwinger-regularized propagator; we therefore observe that the effective cutoff induced by the stochastic RG transformation is
\BE
\mLam_t^{-2} = \mLam_0^{-2} + 2t, \quad \mrm{or} \quad \mLam_t = \frac{\mLam_0}{\sqrt{1+ 2\tau}},
\EE
where the dimensionless flow time $\tau = \mLam_0^2 t$ has been introduced. We will take another look at the effective correlation functions and the function $A_t$ in the next section.

We can make sense of the odd-looking factors of $f_t$ and $B_t$ that appear in the effective action as follows. First, the additive $h_t$ in the propagator $\mDelt_0^{(t)}$ acts as a sliding IR cutoff for the \emph{bare} theory, since\footnote{We choose to consider the mass term in $S_0$ as part of the interaction $V_0(\phi)$ from now on.}
\BE
\lim_{p\to 0} h_t(p) = \frac{1}{t},
\EE
which means that as $t$ increases, more of the bare field modes get integrated out. For example, in the case of $\phi^4_d$ theory (discussed in more detail in the next subsection), the momentum-independent part of the 1-loop contribution to the amputated effective 4-point vertex in $W^{(t)}_0$ is proportional to
\BE
\int_{\mathbb{R}^d} \frac{\dd^d k}{(2\pi)^d} [\mDelt^{(t)}_{0}(k)]^2 = \int_{\mathbb{R}^d} \frac{\dd^d k}{(2\pi)^d} \; \frac{\me^{-2 k^2 a_0^2}}{\big(k^2 + h_t(k) \big)^2} \; .
\EE
We observe that the presence of $h_t$ in the denominator, combined with the multiplicative bare cutoff function, effectively restricts the domain of integration to $\| p \| \in [\mLam_t, \mLam_0]$, similarly to what one would have found in a standard (sharp) high-mode elimination RG step $\mLam_0 \to \mLam$, where the domain of the integral would be $\| p \| \in [\mLam, \mLam_0]$ (see \cite{Kopietz:2010zz} for details). Next, note that the argument $B_t f_t \phi$ of $W^{(t)}_0$ in $S_t$ implies that the $[W^{(t)}_0]^{(n)}$ vertices are mutiplied by a factor of
\BE
B_t(p) f_t(p) = K_{0}^{-1}(p) \frac{2 \omega(p) f_t(p)}{1-f_t^2(p)}
\EE
for each factor of $\phi(p)$. Since the vertices $[W^{(t)}_0]^{(n)}$ are connected $n-$point functions, which have $n$ factors of external propagators $\mDelt_0^{(t)}(p_i) \propto K_0(p_i)$ attached to them, we see that the effective vertices decay like $f_t(p_i) = \me^{-p^2_i t}$ and therefore strongly suppress the $\| p_i \| \gg \mLam_t$ contribution of the $n-$point functions. Moreover, the leading momentum behavior of the products of $B_t f_t$ with $\mDelt^{(t)}_0$ demonstrates that they are, in a sense, \emph{amputated},
\BE
B_t(p) f_t(p) \mDelt^{(t)}_0(p) = 1 - \frac{1}{2} (p^2 t)^2 + O(p^8 t^4),
\EE
in a manner similar to what one finds with sharp high-mode elimination. Thus, in sum, the effective vertices are amputated connected $n-$point functions to leading order in external momenta,  which are heavily damped in the UV $(\| p \| \gg \mLam_t)$, and whose loop corrections effectively involve domains of integration $\| p \| \in [\mLam_t, \mLam_0]$. It is also noteworthy that the external momemtum dependence implied by the amputation formula above goes like powers of $p^2 / \mLam_t^2$, for $\mLam_t^{-2} \gg \mLam_0^{-2}$, as one expects from the general philosophy of effective field theory.

\paragraph{Flow of the couplings and rescaling.} At first sight it appears that the effective action, written as an integration against the bare density, eq. (\ref{eff_action}), has a trivial infinite flow time limit, as
\BE
\lim_{t \to \infty} \rho_t(\phi) = \big[\det 2 \pi B_\infty \big]^{\frac{1}{2}} \exp\Big[- \frac{1}{2} \big(\phi, B_\infty \phi\big)\Big],
\EE
where $B_\infty(p) = 2 K_{0}^{-1}(p) \omega(p)$, due to the exponential decay of $f_t$. Indeed, it is well-known that the Ornstein-Uhlenbeck process has a gaussian stationary distribution. It is also well-known that the stationary distribution of a non-linear Langevin equation with drift $\delta \hat S(\phi) / \delta \phi$ is given by the Boltzmann factor $\exp [ - \hat S(\phi)]$. And yet, we know that the stationary distribution of an RG transformation should correspond to a fixed point theory, which in many cases is interacting.

The first feature to note in response to this paradox is that the properties of the drift $\omega$ imply that the zero mode of the bare field is not suppressed (see eq. (\ref{constraint_functional})); only its variance changes. Since the zero-mode theory is not gaussian, in general, the flowed distribution will also have a non-gaussian zero mode effective action, implying that the long-distance physics is non-trivial. This would suggest, however, that the infinite-time degrees of freedom do not propagate. To further clarify the situation, we will look at the flow of the most relevant effective couplings as the RG time $t$ increases, and then we will address the role of rescaling of degrees of freedom.

As an example of where an IR fixed point is known to exist by other means, we choose to analyze $\phi^4_3$ theory perturbatively, treating the mass term also as a perturbation. Denoting the coefficient of $p^2$ in the quadratic part of $S_t(\phi)$ by $c_t$, and the momentum-independent parts of the quadratic and quartic terms, respectively, by $m^2_t, \; \lambda_t$, we find
\begin{align}
c_t &= 1 + O(\lambda_0^2), \\
m^2_t & = m^2_0 + \frac{\lambda_0}{2} I^d_0(t) + O(\lambda_0^2, \lambda_0 m^2_0), \\
\lambda_t & = \lambda_0 - \frac{3\lambda_0^2}{2} C^d_0(t) - 2 \lambda_0^2 t I^d_0(t) + O(\lambda_0^3, \lambda_0 m^2_0),
\end{align}
at 1-loop order, where the loop integrals are given by
\begin{align}
I^d_{0}(t) & = \int_{\mathbb{R}^d} \! \frac{\dd^d p}{(2 \pi)^d} \frac{\me^{-p^2 a_0^2}}{p^2 + h_t(p)} = \Omega_d \int_{\mathbb{R}_+} \! \dd p \; p^{d-3} \me^{-p^2 a_0^2} \tanh p^2 t, \\
C^d_{0}(t) & = \int_{\mathbb{R}^d} \! \frac{\dd^d p}{(2\pi)^d} \; \frac{\me^{-2 p^2 a_0^2}}{\big(p^2 + h_t(p) \big)^2} = \Omega_d \int_{\mathbb{R}_+} \! \dd p \; p^{d-5} \me^{-2 p^2 a_0^2} \tanh^2 p^2 t,
\end{align}
and $\Omega_d = S_{d-1} / (2\pi)^d$. The first integral is superificially divergent, but for $a_0 > 0$, it has a finite $t\to\infty$ limit, and one may compute
\BE
t \frac{\dd}{\dd t} I^d_0(t) = \Omega_d \alpha_1 \; t^{1-d/2} + O(t^{-d/2} a_0^2),
\EE
where $\alpha_1 \approx 0.379064$ for $d=3$. The second integral $C^d_0(t)$ exists even for $a_0=0$, and its time derivative is
\BE
t \frac{\dd}{\dd t} C^d_0(t) = \Omega_d \alpha_2 \; t^{2 - d/2} + O(t^{2 - d/2 - \delta} a_0^{2\delta}),
\EE
where $\delta > 0$ and $\alpha_2 \approx 0.594978$ for $d = 3$.\footnote{Recall that $\phi^4_3$ theory is superrenormalizable, having only two superficially divergent diagrams: the snail and the sunset diagrams.} Hence, to 1-loop order, we find for the derivatives of effective couplings
\begin{align}
t \frac{\dd}{\dd t} m^2_t & = \frac{\lambda_0}{2} \Omega_d \alpha_1 \; t^{1-d/2} + O(t^{-d/2} a_0^2), \\
t \frac{\dd}{\dd t} \lambda_t & = - \lambda_0^2 \Omega_d \big( \smallfrac{3}{2} \alpha_2 + 2 \alpha_1 \big) \; t^{2-d/2} + O(t^{2 - d/2 - \delta} a_0^{2\delta}).
\end{align}
These expressions do not clearly indicate any nontrivial fixed-point behavior at this order in perturbation theory. To proceed further, one must cast the flow equations in terms of rescaled dimensionless quantities, as one usually does to study RG flows. We will find below that such quantities naturally arise after a passive momentum and field redefinition.

Now we introduce dimensionless rescaled variables using the effective scale $\mLam_t$ to give dimension \cite{Morris:1993qb}.  Dimensionless momenta $\bar p$ are defined by setting
\BE
p =: \mLam_t \bar p.
\EE
The kinetic term in the effective action therefore becomes
\BE
\frac{1}{2} \int_{\bar p} \mLam_t^{d + 2} \bar p^2 \phi(\mLam_t \bar p) \phi(-\mLam_t \bar p).
\EE
This motivates a change of field variables $\phi \to \mPhi$, where $\mPhi$ is dimensionless:
\BE \label{field_cov}
\phi(\bar p \mLam_t) =: \mLam_t^{d_\phi} \mPhi(\bar  p),
\EE
with $d_\phi = -d/2-1$ being the canonical mass dimension of $\phi$ in momentum space. After doing so, the kinetic term is of the canonical form
\BE
\frac{1}{2} \int_{\bar p} \bar p^2 \mPhi(\bar p) \mPhi(- \bar p)
\EE
at 1-loop order, while the mass and quartic terms pick up factors of $\mLam_t$ which define dimensionless couplings $r_t, \; u_t$ by
\BE
r_t := \mLam_t^{-2} m^2_t, \qquad u_t := \mLam_t^{d-4} \lambda_t.
\EE
We note that these rescalings are all quite familiar when written in terms of the scale factor
\BE
b_t := \frac{\mLam_0}{\mLam_t} \quad \Rightarrow \quad r_t = b_t^2 \hat m_t^2, \quad u_t = b_t^{4-d} \hat \lambda_t,
\EE
reflecting that the mass and the 4-point coupling are relevant at the gaussian fixed point (hats denote quantities rendered dimensionless with $\mLam_0$).

Next, we compute the RG flow equations which describe how the dimensionless variables change with the flow time $t$. In the expression for the derivatives above, one replaces $m^2_0$ and $\lambda_0$ by $m^2_t$ and $\lambda_t$, valid at this order in perturbation theory. The derivatives of the dimensionless couplings with respect to $b_t$ are then
\begin{align}
b_t \frac{\dd r_t}{\dd b_t} &= 2 r_t + \beta_1 u_t, \\
b_t \frac{\dd u_t}{\dd b_t} &= (4 - d) u_t - \beta_2 u^2_t,
\end{align}
up to terms of order $b^{-2}_t$, since $t = \frac{1}{2} \mLam_t^{-2}(1-b^{-2}_t)$, and where $\beta_1 = 2^{\frac{1}{2}} \Omega_3 \alpha_1, \; \beta_2 = 2^{\frac{1}{2}} \Omega_3 (\frac{3}{2} \alpha_2 + 2 \alpha_1)$ in $d=3$. As $b_t \to \infty$, the second equation has a nontrivial stationary solution $u_*$, and implies a corresponding critical value $r_*$, which for $d = 3$ are given, at 1-loop order, by $u_* \approx 8.46, \; r_* \approx -0.12$. Linearizing about the fixed point and computing the left-eigenvalues $y_i$ of the stability matrix, one finds that $y_2 = 2, \; y_4 = - 1$, which are crude approximations to the precisely-known values $y_2 = 1.58831(76), \; y_4 = -0.845(10)$ at the Wilson-Fisher fixed point \cite{Hasenbusch:1999mw}.\footnote{The values at this loop order from sharp high-mode elimination combined with epsilon expansion in \cite{Kopietz:2010zz} are $y_2 = 1.67, \; y_4 = -1$, which, however, treats the mass non-perturbatively. As a step in that direction, extending the analysis above to include terms of order $ru$ yields modified exponents $y_2 \approx 1.63, \; y_4 \approx -1.33$; we stress that our formalism is not expected to do any better than the epsilon expansion.}

Thus we observe the existence of an IR fixed point in perturbation theory, as we expect in $\phi^4_3$ theory. If we worked to $O(\lambda_0^2)$, we would find, as usual, the necessity of including a wave function renormalization factor $\zeta_t = b_t^{\gamma_\phi}$ to normalize the kinetic term coefficient $c_t :=1$, so that eq. (\ref{field_cov}) is replaced by
\BE
\phi(\bar p \mLam_t) =: \mLam_t^{d_\phi} \zeta_t^{-1} \mPhi(\bar  p) = \mLam_0^{d_\phi} b_t^{-\Delta_\phi} \mPhi(\bar p),
\EE
which modifies the scaling dimension $\Delta_\phi$ of $\phi$ to include an \emph{anomalous} dimension $\gamma_\phi = O(u_t^2)$, which has a non-zero $t\to\infty$ limit.

The existence of an IR fixed point for the dimensionless, rescaled effective action implies that the expectation values of rescaled effective observables
\BE
\langle \mPhi(\bar p_1) \cdots \mPhi(\bar p_n) \rangle_{S_t} = b_t^{n \Delta_\phi} \mLam_0^{-nd_\phi} \langle \phi(p_1) \cdots \phi(p_n) \rangle_{S_t}
\EE
can have nontrivial infinite flow time limits. In terms of the stochastic RG transformation of section 2, this is written as
\BE
\langle \mPhi(\bar p_1) \cdots \mPhi(\bar p_n) \rangle_{S_t} = b_t^{n \Delta_\phi} \mLam_0^{-nd_\phi}  \big\langle \mathbb{E}_{\mu_0} \big[\phi_t(p_1) \cdots \phi_t(p_n) \big] \big\rangle_{S_0}.
\EE
Since the stochastic RG transformation was generated by a linear Langevin equation, it may be surprising to find that by simply rescaling the correlation functions, one can arrive at a non-gaussian stationary distribution of the Fokker-Planck equation. We also note that the quantities $\mLam_0^{-d_\phi} \phi_t$ correspond directly to the dimensionless field variables one would work with on the lattice.

\paragraph{Correlation functions.}
Wilson and Kogut demonstrated a relation between effective $n-$point functions and the bare $n-$point functions in their FRG scheme \cite{Wilson:1973jj}. Recently, the authors of \cite{Sonoda:2019ibh} have noted that this relation is an equivalence between effective correlations and gradient-flowed correlations. In the context of the stochastic approach here, the corresponding relation is given in terms of generators $W(J)$ of connected Green functions by
\BE
W_t(J) = \frac{1}{2}(J,A_t J) + W_0(f_t J),
\EE
where $A_t$ is given by (eq. \ref{Adef}). This relation is simply derived by shifting $\phi' = \phi - f_t \varphi$ in eq. (\ref{eff_action}) and using the definition of the generator,
\BE
\me^{W_t(J)} := \frac{1}{Z_0} \int \! \mathscr{D} \phi \; \me^{-S_t(\phi) + (J,\phi)},
\EE
with $Z_0$ being the free theory partition function \cite{Kopietz:2010zz, ZinnJustin:2002ru}. It follows that the 2-point functions of $S_t$ and $S_0$ are related by
\BE
W^{(2)}_t(x,y) = A_t(x,y) + \int_{x'y'} f_t(x,x') f_t(y,y') W^{(2)}_0(x',y'),
\EE
and higher $n-$points are related by
\BE
W^{(n)}_t(x_1,\dots, x_n) = \int_{\bo x'} f_t(x_1,x_1') \cdots f_t(x_n,x_n') W^{(n)}_0(x_1',\dots,x_n').
\EE
The function $A_t(x,y)$ is determined by the choice of Langevin equation. In the case $\omega(p) = p^2$, for example, one finds an expression in terms of upper incomplete gamma functions
\BE
A_t(z,0) = \frac{1}{8 \pi^{d/2} z^{d-2}} \Big[\Gamma\Big(\frac{d}{2}-1, \frac{z^2}{4 a_t^2}\Big) - \Gamma\Big(\frac{d}{2}-1, \frac{z^2}{4 a_0^2}\Big) \Big],
\EE
where the inverse effective cutoff $a_t = \mLam_t^{-1}$ was used. For large separations $\| z \| \gg a_t$, this quantity decays as a gaussian. The effective propagator is therefore equal to the gradient-flowed propagator asymptotically in $x-y$ (so long as the correlation length $\xi \gg a_t$):
\BE
\langle \phi(x) \phi(y) \rangle_{S_t} \longrightarrow \langle (f_t\varphi)(x)(f_t\varphi)(y) \rangle_{S_0}.
\EE
Note also that if no cutoff function were imposed on the gaussian noise $\eta_t$, there would be a short-distance singularity in $A_t(z,0)$, regardless of whether the bare theory was regulated.

The connected correlators of composite operators also are simply related to their gradient flow counterparts, except we must be careful to define the generators of their $m-$point functions properly. For example, in the case of $\mcal{O} = \phi^2$, the generator of correlators $W_t^{(0,m)}$ is defined by \cite{ZinnJustin:2002ru, Amit:1984ms}
\BE
\me^{W_t(L)} := \frac{1}{Z_0} \int \! \mathscr{D} \phi \; \me^{-S_t(\phi) + \frac{1}{2}(L,\phi^2)}.
\EE
By inserting the definition of $S_t(\phi)$, one may compute the relation between effective and bare generators exactly, as the integrals involved are gaussian. We note, however, that given the simplicity of the Langevin equation, we can just as easily use the explicit solution $\phi_t[\varphi; \eta]$ to compute expectations. For example, the 2-point correlator of the $\phi^2$ composite operator is
\BE
\langle \phi^2(x) \phi^2(y) \rangle_{S_t}^\con = \langle (f_t\varphi)^2(x)(f_t\varphi)^2(y) \rangle_{S_0}^\con +  A_t(x-y)\langle (f_t\varphi)(x)(f_t\varphi)(y) \rangle_{S_0}^\con + 2 A_t(x-y)^2,
\EE
where the connected part of a correlator of local operators $A, \; B$ is defined by
\BE
\langle A(x) B(y) \rangle^\con := \langle A(x) B(y) \rangle - \langle A(x) \rangle \langle B(y) \rangle,
\EE
which again shows the asymptotic equivalence of effective and gradient-flowed quantities.

In sum, what we have found is that the correlation functions of composite operators in the effective theory are equal to the gradient-flowed correlators, up to terms proportional to powers of $A_t(x-y)$, which itself is determined by the drift term $\omega$. Thus, so long as the drift is chosen to imply an exponentially decaying $A_t$, the flowed observables are sufficient to determine the long-distance properties of the effective theory.

\section{Ratio formulae}

The fact that the transition functional $P_t$ satisfies the Fokker-Planck equation implies that observables at finite $t$ satisfy 
\BE
\frac{\del}{\del t} \langle \mcal{O}(\phi) \rangle_{S_t} = \langle \mcal{L} \mcal{O}(\phi) \rangle_{S_t},
\EE
where the \emph{generator} $\mcal{L}$ of the Markov process is a linear differential operator given by
\BE
\mcal{L} = \frac{1}{2} \mrm{tr} \Big( \mSig(\phi,t) \fdelphi \otimes \fdelphi - 2 \mathscr{B}(\phi,t) \otimes \fdelphi \Big).
\EE
For the flow we have been considering, the generator takes the form
\BE
\mcal{L} = \frac{1}{2} \mrm{tr} \Big( K_0 \fdelphi \otimes \fdelphi \Big) - \mrm{tr} \Big( \omega \phi \otimes \fdelphi \Big),
\EE
where $\omega$ is (minus) the laplacian operator. After a small timestep $\epsilon$, then, successive observables are related by
\BE
\langle \mcal{O}(\phi) \rangle_{S_{t+\epsilon}} = \langle \mcal{O}(\phi) \rangle_{S_{t}} + \epsilon \langle \mcal{L} \mcal{O}(\phi) \rangle_{S_t} + O(\epsilon^2).
\EE
Applied to $n-$point functions, the formula reads\footnote{This formula corresponds to the spin-blocking equation
\BE
\langle B_b \varphi (m_1) \cdots B_b \varphi (m_n) \rangle_{S} = \langle \varphi (m_1) \cdots \varphi(m_n) \rangle_{S} + O(\varepsilon / \Delta m),
\EE
where $B_b \varphi(m) = b^{-d}\sum_\varepsilon \varphi(m+\varepsilon)$ is the blocking operator, $\varepsilon \leq b$, and $\Delta m$ stands for the differences $|m_i - m_j| \gg b \; \forall i \neq j$. This follows from the usual correlator scaling relations of \emph{rescaled} spins $\varphi_b(n/b) := b^{\Delta_\phi} (B_b \varphi)(n)$,
\BE
\langle \varphi_b (m_1/b) \cdots \varphi_b (m_n / b) \rangle_{S_b} = b^{n \Delta_\phi} \langle \varphi (m_1) \cdots \varphi(m_n) \rangle_{S} + O(\varepsilon / \Delta m),
\EE
that one finds in textbooks, e.g. \cite{Cardy:1996xt, Amit:1984ms}.
}
\BE \label{npoint_step}
\langle \phi(x_1) \cdots \phi(x_n) \rangle_{S_{t+\epsilon}} = \langle \phi(x_1) \cdots \phi(x_n) \rangle_{S_{t}} + O(\epsilon).
\EE
Writing both sides in terms of the rescaled theory variables, $\phi(x) \propto \mLam_t^{\Delta_\phi} \mPhi(\bar x)$, where the densionless position is defined by $x = \mLam_t^{-1} \bar x$, one finds
\BE
\mLam_{t+\epsilon}^{n\Delta_\phi} \langle \mPhi(\bar x_1) \cdots \mPhi(\bar x_n) \rangle_{S_{t+\epsilon}} = \mLam_t^{n\Delta_\phi} \big[ \langle \mPhi( \bar y_1) \cdots \mPhi(\bar y_n) \rangle_{S_{t}} + O(\epsilon)\big].
\EE
Motivated by the definition of scale changes $b_t = \mLam_0 / \mLam_t$ with respect to the bare scale, we introduce the \emph{relative} scale change $b_\epsilon(t) := b_{t+\epsilon} / b_t = \mLam_t / \mLam_{t+\epsilon}$. Since the rescaled positions at different scales, $\bar x$ and $\bar y$, refer to the \emph{same} dimensionful position $x$ defined at the bare scale (i.e. in units of $a_0 = \mLam_0^{-1}$), it follows that $\bar y = b_\epsilon \bar x$, and we may write the previous formula as
\BE
 \langle \mPhi(\bar x_1) \cdots \mPhi(\bar x_n) \rangle_{S_{t+\epsilon}} = b_\epsilon(t)^{n\Delta_\phi}  \big[\langle \mPhi(b_\epsilon \bar x_1) \cdots \mPhi(b_\epsilon \bar x_n) \rangle_{S_{t}} + O(\epsilon)\big],
\EE
To the extent that we may neglect the $O(\epsilon)$ terms (which we justify in the appendix), we therefore find a familiar RG scaling relation,
\BE \label{phi_ratios}
 \langle \mPhi(\bar x_1) \cdots \mPhi(\bar x_n) \rangle_{S_{t+\epsilon}} \approx b_\epsilon(t)^{n\Delta_\phi} \langle \mPhi(b_\epsilon \bar x_1) \cdots \mPhi(b_\epsilon \bar x_n) \rangle_{S_{t}}.
\EE
This formula is the stochastic RG analogue of a spin-blocked correlator scaling relation.

We may now derive a scaling relation for correlations with (scaling) operator insertions. Consider a 1-parameter family of actions $S_t(\theta) = S_t - \theta \mcal{V}_\mcal{O}$, where $\mcal{V}_\mcal{O}(\mPhi)$ is the volume-integral of a local operator $\mcal{O}[\mPhi(\bar x)]$ (which may be polynomial in $\mPhi$). $n-$point functions of $\mPhi$ with insertions of $\mcal{V}_\mcal{O}$ in the rescaled effective theory may be written as
\BE
\langle \mcal{V}_\mcal{O}(\mPhi) \mPhi(\bar x_1) \cdots \mPhi(\bar x_n) \rangle_{S_t} = \frac{\dd}{\dd \theta} \langle \mPhi(\bar x_1) \cdots \mPhi(\bar x_n) \rangle_{S_t(\theta)} \Big|_{\theta = 0}.
\EE
Now consider a small time step $\epsilon$ as before. If $\mcal{V}_\mcal{O}$ is an operator in the rescaled action whose coefficient transforms as $g_{t+\epsilon} = b_\epsilon^{y_g} g_t$ (obtained by diagonalizing the linearized flow equations for the couplings),  and if we introduce $\theta$ as a perturbation of $S_{t+\epsilon}$, then eq. (\ref{phi_ratios}) together with the previous equation imply
\BE
\langle \mcal{V}_\mcal{O}(\mPhi) \mPhi(\bar x_1) \cdots \mPhi(\bar x_n) \rangle_{S_{t+\epsilon}} \approx b_\epsilon(t)^{n\Delta_\phi - y_g} \langle \mcal{V}_\mcal{O}(\mPhi) \mPhi(\bar y_1) \cdots \mPhi(\bar y_n) \rangle_{S_t}.
\EE
The minus sign before $y_g$ arises because, with respect to $\theta$, which is the coupling at $t+\epsilon$, the previous step would have coupling $b^{-y_g}_\epsilon \theta$. Since $\dd^d \bar y = b_\epsilon^{d} \dd^d \bar x$, one may infer the scaling of insertions of the local operators $\mcal{O}[\mPhi(\bar x)]$:
\BE \label{rescaled_scaling}
\langle \mcal{O}[\mPhi(\bar x)] \mPhi(\bar x_1) \cdots \mPhi(\bar x_n) \rangle_{S_{t+\epsilon}} \approx b_\epsilon(t)^{n\Delta_\phi + \Delta_\mcal{O}} \langle \mcal{O}[\mPhi(b_\epsilon \bar x)] \mPhi( b_\epsilon \bar x_1) \cdots \mPhi(b_\epsilon \bar x_n) \rangle_{S_t}.
\EE
In other words, the scaling operators transform like $\mcal{O}[\mPhi(\bar x)]|_{S_{t+\epsilon}} = b_\epsilon^{\Delta_\mcal{O}} \mcal{O}[\mPhi(\bar y)]|_{S_t}$ inside of expectation values, where $\Delta_\mcal{O} = d - y_g$ is the scaling dimension of $\mcal{O}$.

Writing the rescaled variables in eq. (\ref{rescaled_scaling}) back in terms of $\phi$, and by using the MCRG equivalence between expectations of $\phi$ and $f_t \varphi$, one finds that the factors of $b_\epsilon^{n\Delta_\phi}$ cancel, and the remaining gradient-flowed quantities satisfy a ratio formula:
\BE
\frac{\langle \mcal{O}[f_{t+\epsilon} \varphi(x)] f_{t+\epsilon} \varphi(x_1) \cdots f_{t+\epsilon} \varphi(x_n) \rangle_{S_0}}{\langle \mcal{O}[f_{t} \varphi(x)] f_{t} \varphi(x_1) \cdots f_{t} \varphi(x_n) \rangle_{S_0}} \approx b_\epsilon(t)^{\Delta_\mcal{O} - m\Delta_\phi},
\EE
where $m$ is the number of factors of $\phi$ in $\mcal{O}$. To reiterate, the position arguments on both sides are the \emph{same} physical positions in units of $a_0$. Since the scaling dimension of $\mcal{O}$ can always be written as $\Delta_\mcal{O} = m d_\phi + \gamma_\mcal{O}$, with $d_\phi$ being the \emph{canonical} dimension of $\phi$ in position space, the exponent of $b_\epsilon(t)$ is just $\gamma_\mcal{O} - m\gamma_\phi$.\footnote{If $\mcal{O}$ contains $\ell$ derivatives, then the scaling dimension will be $\Delta_\mcal{O} = m d_\phi + \ell + \gamma_\mcal{O}$.}

\section{Conclusion}

In this work, it has been demonstrated that a class of stochastic processes on field space defined by a linear Langevin equation may be used to define an FRG transformation on the degrees of freedom of a given bare theory. By suitably rescaling the effective degrees of freedom, the Fokker-Planck distribution generated by the Langevin equation can approach a non-gaussian stationary distribution, in accordance with the requirement that an RG transformation should possibly have a nontrivial fixed point, so long as the dynamics of the theory allows it; this was explicitly checked in the case of $\phi^4_3$ theory. By equating the observables of the effective theory with stochastic observables, a form of MCRG for exact RG has been established, implying the possibility of directly simulating such effective theories on the lattice by supplementing an ensemble average by the integration of a Langevin equation. By further analyzing the properties of effective theory observables, it was found that gradient-flowed correlators are asymptotically (in operator-separation) equal to their effective theory counterparts, and therefore the sufficiency of measuring gradient-flowed quantities at large distances to determine long-distance properties of the effective theory was established. Lastly, the Markov property of the stochastic process was used to derive a scaling formula for the ratios of gradient-flow observables with composite operator insertions, thereby implying a means of measuring anomalous dimensions of such operators.

The utility of the relation of stochastic RG to gradient flow lies in its avoidance of the necessity of a full Langevin equation simulation, subject to the requirement that the observables one measures are correlators at large distances. At short distances the presence of additive contributions from the function $A_t$ imply that the gradient flow observables are not sufficient, and one should properly simulate the full transformation. The fact that the stochastic RG transformation is continuous, together with the observation that the effective action flows towards an IRFP, then suggests the possibility of a continuous counterpart to the equations proposed in \cite{Swendsen:1979gn} which allowed for the measurement of the discrete RG stability matrix using correlations between volume-averaged operators in the action. Work in this direction is underway.

\paragraph{Acknowledgements.} The author thanks Anna Hasenfratz, Ethan Neil, Tom DeGrand, and Masafumi Fukuma for useful discussions and suggestions. This work has been supported by the U. S. Department of Energy under grant number DE-SC0010005.

\appendix

\section{The generator $\mcal{L}$}

Here we demonstrate that the $O(\epsilon)$ terms in eq. (\ref{npoint_step}) decay like a gaussian at large distances. For the $n-$point functions of $\phi$, the action of $\mcal{L}$ yields
\begin{align} \label{L_action1}
\langle \mcal{L}[\phi(x_1) \cdots \phi(x_n)] \rangle_{S_t} & = \sum_{i \neq j}^n K_0(x_i - x_j) \langle \phi(x_1) \cdots \hat \phi(x_i) \cdots \hat \phi(x_j) \cdots \phi(x_n) \rangle_{S_t} \\ 
& + \sum_{i=1}^n \langle \phi(x_1) \cdots \Delta_i \phi(x_i) \cdots \phi(x_n) \rangle_{S_t}, \nonumber
\end{align}
where $\hat \phi(x_i)$ means that $\phi(x_i)$ is absent from the product, and $\Delta_i$ is the laplacian with respect to $x_i$. The kernel $K_0(x_i - x_j)$ decays as a gaussian for $\| x_i - x_j \| \gg a_0$. From the MCRG equivalence, both expectation values may be written in terms of gradient-flow $n-$point functions, up to terms involving $A_t(x_i - x_j)$, which decay like $\me^{-x_{ij}^2 \mLam^2_t}$. For the gradient flow terms, note that an insertion of $\Delta_i (f_t \varphi)(x_i)$ satisfies the heat equation
\BE
\Delta_i (f_t \varphi)(x_i) = \del_t (f_t \varphi)(x_i),
\EE
and therefore the second sum in eq. (\ref{L_action1}) may be written as
\BE
\sum_{i=1}^n \langle (f_t \varphi)(x_1) \cdots \Delta_i (f_t \varphi)(x_i) \cdots (f_t \varphi)(x_n) \rangle_{S_0} = \frac{\del}{\del t} \langle (f_t \varphi)(x_1) \cdots (f_t \varphi)(x_n) \rangle_{S_0}.
\EE
The right-hand side typically decays fast at large distances. For example, in the $d=3$ massless gaussian model, the flowed 2-point function is given by
\BE
\langle (f_t \varphi)(x) (f_t \varphi)(y) \rangle^\mrm{g}_{S_t} = \frac{\mrm{erf}(z \mLam_t/2)}{4\pi z},
\EE
and its time derivative decays at large distances like $\me^{-z^2 \mLam_t^2/4}$. This is a reflection of the fact that gradient flow does not suppress the zero-mode of $\varphi$. We therefore expect it to hold also in the case of interacting theories at criticality, where there are no other scales for $t$ to couple to at large distances.

\bibliographystyle{JHEP}
\bibliography{RG_GF}

\end{document}